\documentstyle[preprint,aps,epsf]{revtex}   
\begin{document}

\title{\bf The effect of monomer evaporation on a simple 
model of submonolayer growth}

\author{Pablo Jensen (a), Hern\'an Larralde (b) and Alberto Pimpinelli (c)}

\address{(a) D\'epartement de Physique des Mat\'eriaux, Universit\'e Claude
Bernard Lyon-1, 69622 Villeurbanne C\'edex, France;\\
(b) Instituto
de F\'{\i}sica, Lab. de Cuernavaca, Apdo. Postal 48-3, C.P. 62251, Cuernavaca, 
Mor., Mexico;\\
(c) Institut Laue-Langevin, B.P. 156X, F-38042 Grenoble Cedex 9,
 France}

\maketitle

\begin{abstract}
We present a model for thin film growth by particle deposition that
takes into account the possible evaporation of the particles deposited
on the surface.  Our model focuses on the formation of two-dimensional
structures. We find that the presence of evaporation can dramatically
affect the growth kinetics of the film, and can give rise to regimes
characterized by different ``growth'' exponents and island size
distributions. Our results are obtained by extensive computer
simulations as well as through a simple scaling approach and the
analysis of rate equations describing the system. We carefully discuss the
relationship of our model with previous studies of the same physical 
situation, and we show that our analysis is more general.
\end{abstract}


\newpage
The growth of nanostructures and thin films prepared by atomic
deposition is recognized to be both a standard and a promising way to
prepare new materials \cite{doit,laszlo,vipi,larecherche}. Several
models \cite{villain,tang,model,boston,bales} have led to a good
understanding of the growth properties in the simplest cases, when
only a limited number of physical ingredients are included in the
simulations: Deposition, Diffusion and Aggregation as in the DDA model
\cite{boston}. A drawback of this conceptual simplicity is that the
range of experimental situations accurately described by these models
is limited. It is actually limited to (beautifully) artificial
experimental setups where great care is taken to avoid complications
(contamination, surface defects, etc. see for example
\cite{mo,roder,noneq}).  Clearly, many (technologically) interesting
experimental situations are much more complex. Progress towards their
understanding demands the inclusion of processes that have been left
out of the first models. For example, including reversible aggregation
\cite{ratsch} allows to understand the saturation of island density
before island-island coalescence and produces compact islands. In this
paper we show the effects of including {\it evaporation} of the atoms
from the surface.

Evaporation, i.e. the possibility of desorption of adatoms from the
surface, is a feature that should be observed for any system at high
enough temperatures. In this sense, it is a phenomenon that is as
general as the rest of the ingredients of recent models of film
growth, and is capable of completely changing the quantitative
behaviour of the system (scaling dependence of island density on the 
deposition parameters, island size distributions \ldots).  
Moreover, the effects of evaporation have
already been studied experimentally \cite{robins,stowell,henry}.  Thin
film growth models which include evaporation have already been studied
using a mathematical analysis of rate equations
\cite{vipi,stowell,venables73,venables84,stoyanov}. Computer
simulations of such models, aiming a quantitative analysis of scaling relations or
island size distributions have, to our knowledge, never been carried
out. The point is that computer simulations represent an "exact" way
of reproducing the growth, in the sense that they avoid the mean-field
approximations of mathematical rate-equations approaches
\cite{bales,japan,mrs}.  We find that a careful consideration of the
downward transport of monomers from islands leads to results that
significantly differ from previous studies
\cite{stowell,venables73,venables84,stoyanov}.

Evaporation is of course not the only process that needs to be added
in order to better describe most of the film growth experiments. Other
candidates include interlayer transport, particle dissociation (if the
incident particles are molecules or clusters rather than atoms) or
intricate chemical interactions between adatoms and the surface (for a
taste of the real growth world, see for example
refs. \cite{mrs95,surfconf}). These effects are less pervasive than
evaporation, but nevertheless, should eventually be considered, keyed
to specific experimental systems.

The paper is organized as follows. Section I briefly presents the
model and discusses some of its approximations. Then, in section II,
we give a rapid overview of the growth of the films when evaporation
is taken into account, and try to give a {\it physical intuition} for
what is going on. After this, section III presents a simple {\it
scaling} approach which gives the behaviour of the saturation island
density as a function of the parameters. This scaling approach is
completed in section IV by a more rigorous approach based on a careful
analysis of the {\it rate equations} of the system.  Section V
confirms the preceding results by extensive {\it computer simulations}
of the model. Finally, in section VI, we discuss our results and the
differences with the previous approaches 
\cite{stowell,venables73,venables84,stoyanov}.

\section{Presentation of the model}

In this work we will describe the properties of a still
oversimplified submonolayer thin film growth model which includes the
four most important physical ingredients of these systems:

{(1)} {\it Deposition}.  We will assume that atoms are deposited at
randomly-chosen positions of the surface at a flux $F$ per unit
surface per unit time.  

{(2)} {\it Diffusion}.  Isolated adatoms can move in a random
direction by one diameter, or one lattice spacing, which we will take
as our unit length.  We denote by $\tau$ the characteristic time
between diffusion steps.

{(3)} {\it Evaporation}. Isolated adatoms can evaporate off the
surface at a constant rate. We denote by $\tau_e$ the mean lifetime of
a free adatom on the surface.  As an approximation, we will assume
that the these desorbed atoms do not return to the surface. It is
useful to define $X_S=\sqrt{D\tau_e}$ the mean diffusion length on the
substrate before desorption.

{(4)} {\it Aggregation.} If two adatoms come to occupy neighboring
sites, they stick to form an island. We will mainly consider the case
of irreversible aggregation, but a more general analysis of reversible
aggregation, assuming for simplicity the existence of a critical size
$i*$ below which islands tend to dissociate is given. In any case,
islands are assumed to be immobile and do not evaporate.
 
In the following, we call {\it particles} or {\it adatoms} the
isolated atoms (or monomers) that are deposited on the surface, and
{\it islands} a set of connected particles (thus excluding the
monomers).

Some remarks on the assumptions of this simple model regarding its
connection to the experiments are now addressed.

(a) {\it Second layer}--- When a particle ``falls'' on top of an
island, we assume that the particle deposited on the second layer
moves and evaporates essentially as any other particle, except that
its diffusion constant and evaporation rate now correspond to the
process ocurring on a substrate of the same element as the deposited
particles. Practically, the mean diffusion length on the {\it island}
before desorption, defined as $X_S^*=\sqrt{D^*\tau_e^*}$, where the
$^*$ indicate the values on the island, can be different from
$X_S$. As the simplest scenario, we assume that when the particle
reaches the border of the island, it immediately jumps down and
increases the area of the island.  We discuss when this effect can be
ignored without affecting the scaling results, which leads to
applications to situations in which downward transport of monomers
from islands is highly improbable. This could be the case when: ({\it
i\/}) there is a barrier \cite{schwoebel} at the edges of the (first
layer) clusters which prevents single particles from ``falling'' on
the substrate and/or ({\it ii\/}) Particle diffusion on the second
layer is much smaller than diffusion on the substrate.

(b) {\it Island diffusion}---We neglect in this model the possibility
for dimers, trimers or larger islands to diffuse on the
substrate. Island diffusion has been observed in experiments
\cite{cludifexp} and molecular-dynamics simulations
\cite{depristo}. The effects of island diffusion on the growth of thin
films {\it without evaporation} have been addressed in
Refs. \cite{boston,japan,mrs,villain,metiu,barteltdimer}. While these
effects are significant since they change the growth exponents, they
do not modify fundamentally the growth mechanisms. We assume that the
same is true when evaporation is included.

We study here the {\it first} stages of the growth, roughly until the
number of islands on the substrate saturates. The reason is that it is
in this stage that such a simplified model can be of some help to
experimentalists who want to understand the microscopic processes
present in their experiments. These fundamental microscopic processes
are most easily detected in the first stages of the growth, since in
the subsequent stages additional processes can be involved (additional
diffusion paths, interlayer transport, geometrical details of the
lattice\ldots).

\section{Qualitative description} 

Before going into the details of the calculations and their
confirmation by computer simulations, we present a simple picture of
the growth mechanisms of the submonolayer structures under
consideration. For simplicity, we only consider here the case $i^*=1$,
i.e. irreversible aggregation and we neglect the particles deposited
on top of the islands (despite the apparent lack of generality of this
last hypothesis, the results obtained here are quite general, as we
show in section IV).

The qualitative evolution of the system is essentially as follows. The
system initiates as a clean empty surface. Monomers are then deposited
at a constant rate on the surface and are allowed to diffuse and/or
desorb (evaporate). When two (or more) monomers meet, they aggregate
irreversibly to form a static island.  As more of these encounters
occur, the island density increases with time and islands become
larger by capturing adatoms. At some point, islands are so large that
they start touching (coalescing), and monomers are rapidly
captured. These two effects lead to a saturation in the number of
islands. Interestingly, the saturation is attained when the surface
coverage reaches a value close to .15, independently of the parameter
values. This last point is discussed in detail below. We now turn to a
more detailed discussion of the evolution of the systems in two
limiting cases : complete condensation (evaporation is negligible) and
strong evaporation.

\begin{figure}
\centerline{
\hbox{(a)
\epsfxsize=8cm
\epsfbox{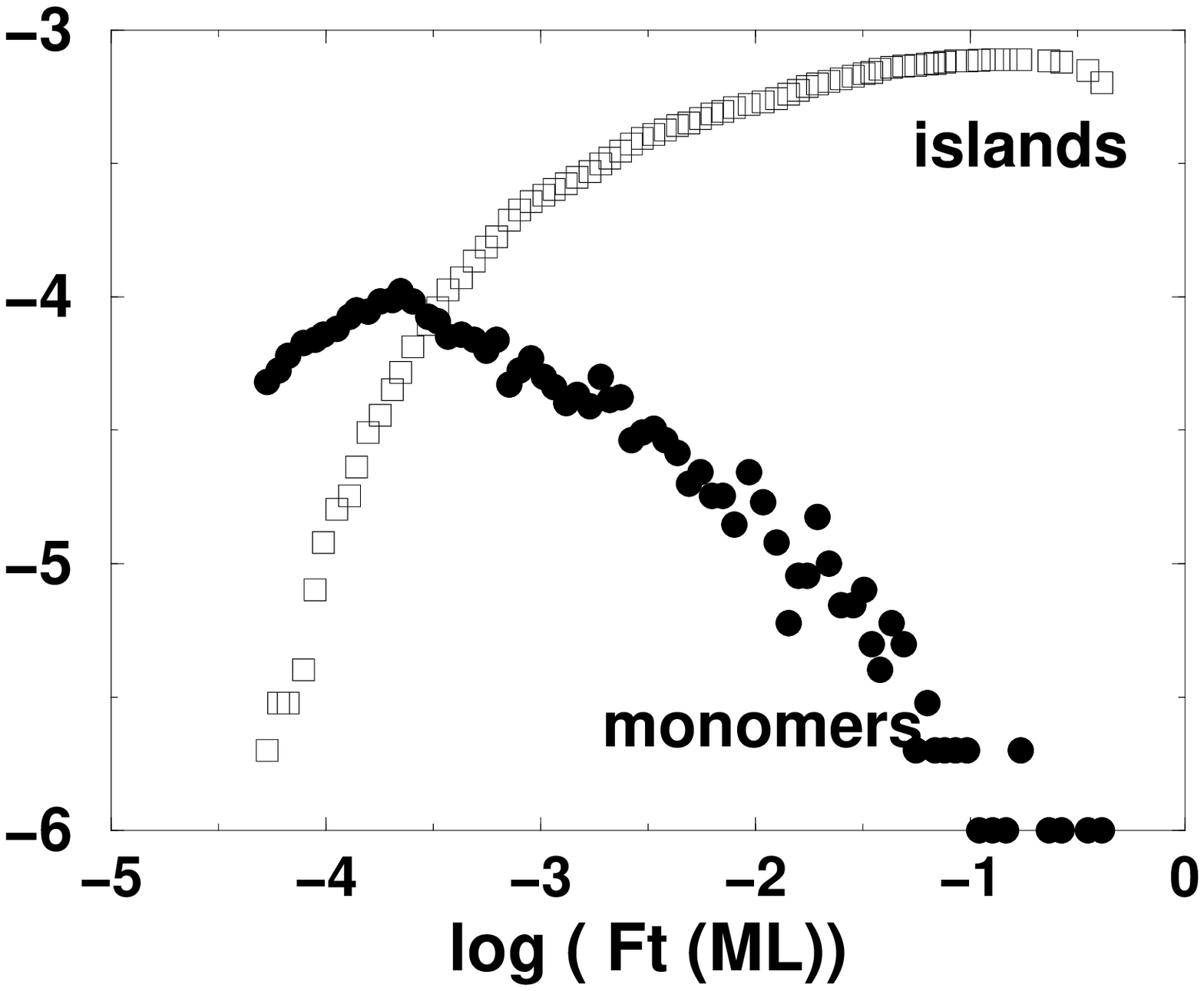}
\hspace{.5cm}(b)
\epsfxsize=8cm
\epsfbox{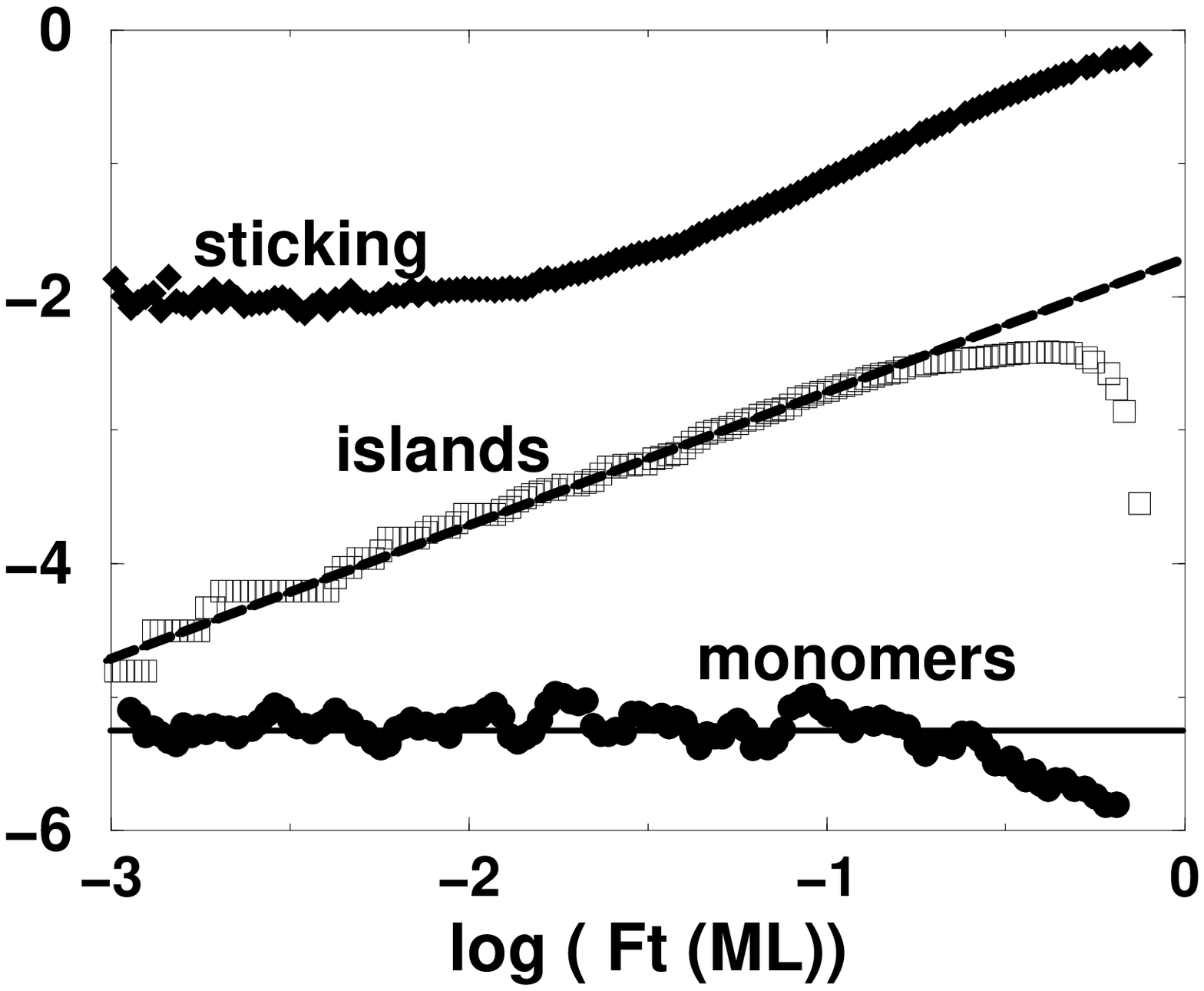}}
}
\caption{
{\it Evolution of the monomer and island densities as a function
of the deposition time (in monolayers) for different limits : (a)
complete condensation, $F=10^{-8}, \ \tau_e=10^{10} \ (\tau=1)$. These
values mean : $X_S = 10^5$ and $\ell_{CC} = 22$ (b) important
evaporation, $F=10^{-8}, \ \tau_e=600 \ (\tau=1)$. These values mean :
$X_S = 25$ and $\ell_{CC} = 22$.  In (b) the "sticking" curve
represents the total number of particles actually present on the
surface (the coverage) divided by the total number of particles sent
on the surface (F t). It would be 1 for the complete condensation
case, neglecting the monomers that are deposited on top of the
islands. The island density curve has been shifted vertically by
(+1). The solid line represents the constant value expected for the
monomer concentration, while the dashed line corresponds to the {\it linear} increase of the island density (see text).}
}
\label{den}
\end{figure}

First consider the situation where evaporation is negligible. This
means that atoms deposited on the surface almost never evaporate
before aggregating (after this, they are safe since islands do not
evaporate). This situation can be expected to happen when $X_S \gg
\ell$ where $X_S=\sqrt{D\tau_e}$ is the adatom diffusion length on the
substrate before desorption and $\ell$ is the typical distance between
islands. Fig \ \ref{den}a shows the evolution of the monomer and
island densities as a function of deposition time. We see that the
monomer density rapidly grows, leading to a rapid increase of island
density by monomer-monomer encounter on the surface. This goes on
until the islands occupy a significant fraction of the surface,
roughly 1\%. Then, islands capture rapidly the monomers, whose density
decreases. As a consequence, it becomes less probable to create more
islands, and we see that their number increases more slowly. When the
coverage reaches a value close to 15\%, coalescence will start to
decrease the number of islands. The maximum number of islands
$N_{max}$ is thus reached for coverages around 15\%.  Concerning the
dependence of $N_{max}$ as a function of the model parameters, the DDA
and related models have shown that the maximum number of islands per
unit area formed on the surface scales as $N_{max} \simeq
(F/D)^{1/3}$, or $\ell_{CC} \simeq (F/D)^{-1/6}$
\cite{villain,venables84,stoyanov} where $CC$ stands for Complete
Condensation.  These values are independent of $\tau_e$ since
evaporation is not significant.

\begin{figure}
\centerline{
\hbox{
\epsfxsize=6cm
\epsfbox{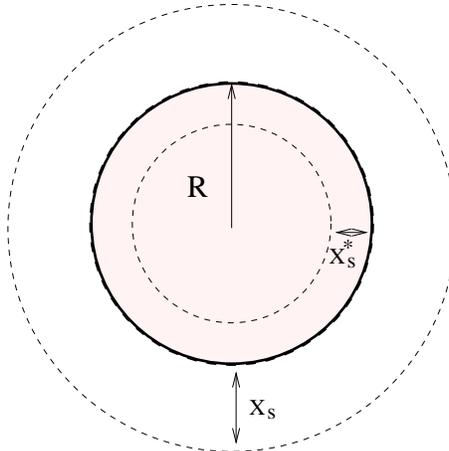}}
}
\caption{
 Schematic capture zone for an island in presence of atom
evaporation. $X_S^*$ stands for the mean length before desorption for
an atom diffusing {\it on top} of an island, whereas $X_S$ corresponds
to the same parameter on the substrate (see text). The sizes of the
capture zones will be justified below, in section IV.}
\label{capture}
\end{figure}

When $\tau_e$ decreases (i.e. as the evaporation rate increases),
$X_S$ decreases and eventually becomes smaller than $\ell_{CC}$.  In
this regime, evaporation is going to significantly alter the growth
dynamics, as shown in Fig \ \ref{den}b. The main point is that now the
monomer density becomes roughly a {\it constant}, since it is now
mainly determined by the balancing of deposition and evaporation. As
expected, the constant concentration equals $F \tau_e$, as shown by
the solid line. Then the number of islands increases linearly with
time (the island creation rate is roughly proportional to the square
monomer concentration). We also notice that only a small fraction
(1/100) of the monomers do effectively remain on the substrate, as
shown by the low sticking coefficient value at early times (the
sticking coefficient is the ratio of particles on the substrate (the
coverage) over the the total number of particles sent on the surface
(Ft)).  This can be understood by noting that the islands grow by
capturing only the monomers that are deposited within their "capture
zone" (comprised between two circles of radius $R$ and $R+X_S$ if we
neglect $X_S^*$, see Figure \ \ref{capture}).  The other monomers
evaporate before reaching the islands.  As in the case of complete
condensation, when the islands occupy a significant fraction of the
surface, they capture rapidly the monomers. This has two effects : the
monomer density starts to decrease, and the sticking coefficient
starts to increase. Shortly after, the island density saturates and
starts to decrease because of island-island coalescence.

If $\tau_e$ is further decreased (precisely $\tau_e \leq \tau$
i.e. the time a particle remains on the surface is less than the time
it needs to move), then, clearly, diffusion plays no role ("$X_S \leq
1$"). In this situation, islands are formed by {\it direct
impingement} of incident atoms as first neighbors of adatoms, and grow
by direct impingement of adatoms on the island boundary. This
situation, although apparently uncommon, is not physically impossible
and it also allows us to test our predictions over a larger range of
parameters.

\section{Scaling arguments}

In this section we present simple scaling arguments that allow to find
the dependence of the maximum island density $N_{max}$ as a function
of the deposition parameters (Flux F, Diffusion time $\tau$ and
Evaporation time $\tau_e$).  These arguments were originally
formulated in \cite{villain} for the special case of deposition
without evaporation on a high-symmetry terrace. Here, the argument is
extended to the case of non-negligible evaporation. We recall that the
atomic size is taken as the length unit. For simplicity, we neglect in
this section the effects of deposition on the second layer : they will
be studied in great detail in the next section. We will show there
that the regime studied in this section corresponds to a large range
of physical situations.

The first stage of the argument requires the determination of the
nucleation rate per unit surface and time, $1/\tau_{\rm nuc}$. A
nucleation event takes place when two adatoms meet. This happens with
a probability per unit time $D\rho^2$, where $D=1/(4 \tau)$ is the
adatom diffusion constant, and $\rho$ the adatom density. Thus,
\begin{equation}
\label{nuc1}
\frac{1}{\tau_{\rm nuc}} \approx D \rho^2
\end{equation}

Another, independent equation can be written down to relate the
nucleation rate and the island density $N$. It states that in the area
$\ell_s^2 = 1/N$ occupied by an island, only one (on average)
nucleation event takes place, {\it during the time $t_c$ needed for
the growing islands to come into contact}. Thus,
\begin{equation}
\label{nuc2}
\frac{1}{\tau_{\rm nuc}} \approx \frac{N}{t_c} \;.
\end{equation}

The time $t_c$ is readily computed by knowing the growth velocity of
an island, which in turn requires the knowledge of the adatom
density. We consider in the following the situations of interest for
this paper.

\subsection{No evaporation, compact 2-D islands}

The adatom density results in this case from a balance between
deposition at a rate $F$ and capture by the stable islands at a rate
$D \rho N$, so that
\begin{equation}
\label{n1}
\rho \approx  F/(DN) \approx F\ell_s^2/D \
\end{equation} 

The area of an island of linear size $R$ increases by capture of all
adatoms falling in an area $\ell_s^2$, so that $R^2\approx F\ell_s^2
t$. At $t=t_c$, $R\approx\ell_s$, and thus $t_c\approx 1/F$. Using
(\ref{nuc1}) and (\ref{nuc2}) one finds $D \rho^{2} \approx FN$. From
(\ref{n1}), $N\approx F\ell_s^4/D$, or
\cite{villain,venables84,stoyanov}
\begin{equation}
N\approx \left( \frac FD\right)^{1/3}\;.
\label{scpinev}
\end{equation}

The value $1/3$ corresponds to the special case $i^*=1$ of the general formula
$i^*/(2+i^*)$ for the exponent \cite{villain,venables84,stoyanov}.

\subsection{Strong evaporation}
 
Strong evaporation means the adatoms are more likely to disappear due
to desorption, with probability $1/\tau_e$ per unit time and site,
than to be captured by an island. Therefore, the adatom diffusion
length before desorption, $X_S=\sqrt{D\tau_e}$, is shorter than the
average island-island distance, $\ell_s$. In this case, the adatom
density results from a balance between deposition and desorption at a
rate $\rho/\tau_e$, so that
\begin{equation}
\label{n2}
\rho \approx F\tau_e\approx FX_S^2/D .
\end{equation} 

A growing island is only able to capture the adatoms falling at a
distance smaller than the adatom diffusion length before desorption,
$X_S$ (see Figure \ \ref{capture}).  If $X_S$ is smaller than the
island size $R$, the island will capture the adatoms falling inside an
annulus of width $\approx X_S$ around its border, as well as those
directly impinging on its edge. Since the area of the annulus is
$\approx RX_S$ if $R \gg X_S$, at time $t$ one has $R\approx
F(1+X_S)t$.  At $t=t_c$, $R\approx\ell_s$, and thus $t_c\approx
\ell_s/[F(1+X_S)]\approx 1/[FN^{1/2}(1+X_S)]$. Direct impingement is
also important for nucleation, so that (\ref{nuc1}) becomes
\begin{equation}
\label{nuc3}
(F+D \rho)\rho\approx F(1+X_S^2)\rho\approx
F^2\tau_e(1+X_S)^2\
\end{equation}
where we also used (\ref{n2}).

From (\ref{nuc2}) and (\ref{nuc3}), one finds $F^2\tau_e(1+X_S)^2
\approx F(1+X_S)N^{3/2}$, or
\begin{equation}
N\approx \left(F\tau_e\right)^{2/3}\left(1+X_S\right)^{2/3}\;.
\label{scpiev}
\end{equation}

For a comparison with previous results, it may be worth it to rederive
(\ref{scpiev}) for any value of $i^*$. The nucleation rate reads, for
any $i^*$, $(F+D\rho)N_{i^*}$, where $N_{i^*}$ is the density of
critical nuclei (clusters of size $i^*$). Following
Refs. \cite{venables84,stoyanov}, we assume that $N_{i^*}$ satisfies
Walton's relation $N_{i^*}\sim \rho^{i^*}$.  Inserting into
(\ref{nuc3}) yields $1/\tau_{\rm nuc}\approx
F(F\tau_e)^{i^*}(1+X_S)^2$. Finally, using (\ref{nuc2}) yields
$F(F\tau_e)^{i^*}(1+X_S)^2\approx F(1+X_S)N^{3/2}$, or
\begin{equation}
N\approx \left(F\tau_e\right)^{2i^*/3}\left(1+X_S\right)^{2/3}\;.
\end{equation}

Note that the validity of Walton's relation in the presence of strong
desorption is highly hypothetical. It is a point which we reserve for
future work. In the rest of the paper, we only consider the case
$i^*=1$. In section IV, the results for $i^*=1$ are derived by a more
rigorous rate-equations \cite{zinsmeister} approach.

\subsection{Crossover scaling}

We will derive here an approximate analytic expression for the
crossover scaling function connecting the complete condensation,
diffusion and direct impingement regimes described previously. In
other words, we will compute the maximum island density $N_{max}$ as a
function of $F$, $\tau$ and $\tau_e$. We will then show that, if we
measure all lengths in terms of $X_S$, $N_{max}(F,\tau,\tau_e)$ is a
function of its arguments only through a special combination:
$X_S^2N_{max}=g_{evap}(X_S^2N_{evap})$, where
$N_{evap}=\left[(1+X_S)F\tau_e\right]^{2/3}$, and $g_{evap}(x)$
satisfies
\begin{equation}
\label{gevap}
g_{evap}(x)\sim\cases{{x^{1/2}} & {\rm for} ~$x\to\infty$,\cr\cr
                   {{x}} & {\rm for}~ $x\to 0$.\cr}
\end{equation}

To do this, we assume that the islands are large enough with respect
to the atomic size, that we can neglect the curvature of their
boundary. Then, the adatom density in the region between two islands
whose edges are at a distance $\ell$, obeys the equation
\begin{equation}
\label{bcf}
\dot \rho = F+D\nabla^2\rho-\frac \rho{\tau_e} \;,
\end{equation}
with the boundary conditions $\rho(\pm\ell/2)=0$ (the origin being midway
between the islands).

In the quasi-stationary approximation, $\dot \rho \approx 0$, and equation
(\ref{bcf}) can be solved. The solution reads
\begin{equation}
\label{sol1}
\rho(x) = F\tau\left[ 1-\frac{\cosh(\kappa x)}{\cosh(\kappa
\ell/2)}\right] \;,
\end{equation}
where $\kappa=1/X_S$

This formula is needed to compute the nucleation rate $D\rho^2$. The
latter is a mean field quantity, independent of $x$. Letting thus
$x=0$ in (\ref{sol1})--since the highest nucleation probability is at
the terrace centre, given the symmetry of our problem--one finds
\begin{equation}
\label{sol2}
\rho = F\tau\left[ 1-\frac{1}{\cosh(\kappa \ell/2)}\right] =2F\tau
\frac{\sinh^2(\kappa \ell/4)}{\cosh(\kappa \ell/2)}\;,
\end{equation}
where we used the identity $\cosh(x)-1=2\sinh^2(x/2)$.

The next task is the determination of the total island density
$N$. Its time variation is simple: $N$ increases each time a new
island is nucleated, so that $(\dot N)_1 = D\rho^2$.  On the other
hand, $N$ decreases when two islands touch and coalesce. Following
previous authors \cite{venables73,stoyanov}, we write $(\dot N)_2 =
-(d {\cal A}/d t) N^2$, where ${\cal A}\approx R^2$ is the average
area of an island of linear size $R$.  This means that coalescence
results from binary encounters of immobile islands, whose area
increases at a rate $d {\cal A}/d t$. Collecting $(\dot N)_1$ and
$(\dot N)_2$ yields
\begin{equation}
\label{isl1}
\dot N = D\rho^2-\frac{d {\cal A}}{d t} N^2 \;.
\end{equation}

By definition, $N_{max}$, which is what we are interested in,
satisfies $\dot N=0$.  This means that the maximum of the island
density can be found by balancing the nucleation rate against the
coalescence rate, and that the balance is reached when the island size
$R\approx\ell$. One can thus write
\begin{equation}
\label{isl2}
  D\rho^2\approx \left(\frac{d {\cal A}}{d t}\right)_{R\approx\ell} N_{max}^2
\;.
\end{equation}

The growth rate of an island can be computed assuming that its radius
is large with respect to the atomic size. One can then treat the
island edges as straight steps, which yields
\begin{equation}
\frac {d R}{d t} \approx  FX_S \tanh(\kappa \ell/2) \;.
\end{equation}

Since $d {\cal A}/d t\approx Rd R/d t$, one finds
\begin{equation}
\label{gr}
\frac {d {\cal A}}{d t}\vert_{R=\ell} \approx
FX_S\ell\tanh(\kappa\ell/2)
\;.
\end{equation}

Note that the two limits $({d {\cal A}}/{d t})\vert_{R=\ell}\approx
F\ell^2$ and $({d {\cal A}}/{d t})\vert_{R=\ell}\approx F\ell X_S$ are
well reproduced for $\kappa\ell \ll 1$ and for $\kappa\ell \gg 1$,
respectively. In fact, in the limit $X_S\to0$, the area increase is
$FR$, due to direct impingement. In order to interpolate all the way
through to this regime, we finally write
\begin{equation}
\label{da}
\frac {d {\cal A}}{d t}\vert_{R=\ell} \approx
F(1+X_S)\ell\tanh(\kappa\ell/2)
\;.
\end{equation}

In the limit where the diffusion length $X_S$ becomes very small, the
nucleation probability is also determined by direct impingement of a
beam atom on a nucleus, which is described by $F\rho$. Then, the
nucleation term in (\ref{isl2}) reads
\begin{equation}
\label{nn}
(F+D\rho)\rho   \approx F\left[1+  X_S^2\frac{\sinh^2(\kappa
\ell/4)}{\cosh(\kappa\ell/2)} \right]\left[F\tau_e\frac{\sinh^2(\kappa
\ell/4)}{\cosh(\kappa\ell/2)}\right]\;.
\end{equation}
In fact, a simpler approximate expression will be used, which
reproduces the correct limiting behaviour at large and small $X_S$:
\begin{equation}
\label{nnn}
(F+D\rho)\rho   \approx\frac{\left(1+  X_S\right)^2}{\tau_e}
\left[F\tau_e\frac{\sinh^2(\kappa
\ell/4)}{\cosh(\kappa\ell/2)}\right]^2\;.
\end{equation}

 Finally,  (\ref{da}) and (\ref{nnn}) yield
\begin{equation}
\label{sc}
\frac{\left(1+  X_S\right)^2}{\tau_e}
\left[F\tau_e\frac{\sinh^2(\kappa
\ell/4)}{\cosh(\kappa\ell/2)}\right]^{2}\approx
F(1+X_S)\ell\tanh(\kappa\ell/2)  N_{max}^2
\;,
\end{equation}
or, letting $\ell= N_{max}^{-1/2}$,
\begin{equation}
\label{sche1}
{\left(1+  X_S\right)}(F\tau_e)
\approx \left[\frac{\cosh(\kappa  N_{max}^{-1/2}/2)}{\sinh^2(\kappa
  N_{max}^{-1/2}/4)}\right]^{2}
\tanh(\kappa  N_{max}^{-1/2}/2)  N_{max}^{3/2} \;.
\end{equation}

This is the announced crossover scaling formula, in implicit form. It
can be cast in the form:
\begin{equation}
\label{sc1}
{(1+  X_S)}X_S^3(F\tau_e)
\approx \left[\frac{\cosh(\kappa  N_{max}^{-1/2}/2)}{\sinh^2(\kappa
  N_{max}^{-1/2}/4)}\right]^{2}
\tanh(\kappa  N_{max}^{-1/2}/2) (X_S^2  N_{max})^{3/2}\;,
\end{equation}
or,
\begin{equation}
\label{eqsca}
{\left(1+ X_S\right)}X_S^{5}\left(\frac FD\right)
 = f_{evap}(X_S N_{max}^{1/2})\;,
\end{equation}
where
\begin{equation}
\label{function}
f_{evap}(x) =
\left[\frac{\cosh\left(\frac{1}{2x}\right)}{\sinh^2\left(\frac{1}{4x}
\right)}\right]^{2}
\tanh\left(\frac{1}{2x}\right) x^{3}\;.
\end{equation}
Letting $N_{evap} =[(1+  X_S)F\tau_e]^{2/3}$, and inverting $f$, one finds as
promised:
\begin{equation}
X_S^2 N_{max}= g_{evap}(X_S^2N_{evap})\;.
\end{equation}

\section{Rate equations} 

In this section we will study the model taking into account the
particles deposited on top of the islands with their specific
diffusion and evaporation parameters. We will show that for a large
range of parameters, the exponents are the same as those predicted by
the preceding analysis. We use a "rate-equations" approach which has been
shown to give a good description of the submonolayer regime \cite{tang,bales,venables73,stoyanov,zinsmeister}. For the sake of completeness,
we  include a detailed calculation of the ``cross sections'' 
for monomer capture. These ``cross sections'' have also been calculated in \cite{bales,stowell,venables73} for example.

We will not keep careful track of many of the numerical geometric
constants. The rate equation describing the time evolution of the
density $\rho$ of monomers on the surface will be, to lowest relevant
orders in F:
\begin{equation}
{d\rho\over dt}=F(1-\theta)-{\rho\over\tau_e} -F\rho - 
2\sigma_o\rho -\sigma_i N
\label{rho}
\end{equation}
The first term on the right hand size denotes the flux of monomers
onto the island free surface, ($\theta$ is the island coverage
discussed below).  The second term represents the effect of
evaporation, i.e. monomers evaporate after an average time
$\tau_e$. The third term is due to the possibility of losing monomers
by effect of direct impingement of a deposited monomers right beside a
monomer still on the surface to form an island. As discussed above,
this ``direct impingement'' term is usually negligible, and indeed
will turn out to be very small in this particular equation, but the
effect of direct impingement plays a crucial role in the kinetics of
the system in the high evaporation regimes. The last two terms
represent the loss of monomers by aggregation with other monomers and
with islands respectively. The factors $\sigma_o$ and $\sigma_i$ are
the ``cross sections'' for encounters and will be calculated below.

The number $N$ of islands will be given by:
\begin{equation}
{dN\over dt}=F\rho+\sigma_o\rho
\label{islands}
\end{equation}
where the first term represents the formation of islands due to direct
impingement of deposited monomers next to monomers already on the
surface, and the second term accounts for the formation of islands by
the encounter of monomers diffusing on the surface.

For the island coverage $\theta$ i.e. the area covered by all the
islands per unit area, we have:
\begin{equation}
{d\theta\over dt}=2\left[F\rho+\sigma_o\rho\right]
                  +\sigma_i N + JN 
\label{cove}
\end{equation}

The term in brackets represents the increase of coverage due to
formation of islands of size 2 (i.e. formed by two monomers) either by
direct impingement or by monomer-monomer aggregation. The next term
gives the increase of coverage due to the growth of the islands as a
result of monomers aggregating onto them by diffusion, and the last
term represents the growth of the islands due to direct impingement of
deposited monomers onto their boundary, or directly on the island. The
estimation of the value of J is carried out below. The total surface
coverage is given by $\theta+\rho\sim\theta$ except at very short
times.

The next step in the analysis consists in estimating the cross
sections $\sigma_o$ and $\sigma_i$. This is done by evaluating the
diffusive flux of monomers into a single single solitary island
represented by an absorbing disk of radius $R$ centered at the origin
(throughout this analysis $R$ stands for the typical island
radius). The typical radius $R$ of the islands will be taken to be:
\begin{equation}
R\sim \left({\theta\over N}\right)^{1/2}
\end{equation}

The cross sections are evaluated in the quasistatic approximation,
which consists in assuming that $R$ does not vary in time and that the
system is at a steady state.  Thus, we have to solve the equation
\begin{equation}
D \left({\partial^2 P\over \partial r^2} +{1\over r}{\partial P\over
 \partial r}\right)+F-{1\over\tau_e} P({\bf r})=0
\end{equation}
subject to the boundary condition $P(r=R)=0$, where $P({\bf r})$ is
the monomer concentration at position ${\bf r}$ and $D$ is the
diffusion constant of the monomers. The solution is given by:
\begin{equation}
P({\bf r})={F \tau_e}\left(1-{K_0(r/X_S)\over K_0(R/X_S)}\right),
\label{ftaue}
\end{equation}
where $X_S=(D\tau_e)^{1/2}$ can be identified as the characteristic
distance traveled by a monomer before evaporation, and $K_0(x)$ is the
modified Bessel function of order zero. We can rewrite Eq.\
\ref{ftaue} as:
\begin{equation}
P({\bf r})=\rho\left(1-{K_0(r/X_S)\over K_0(R/X_S)}\right),
\end{equation}
where $\rho$ is the monomer density appearing in the rate equations,
so that $P({\bf r})\to \rho$ far from the islands. This, again, is a
mean field way of including the effects of the other islands in the
flux of monomers into the island in consideration. The cross section
$\sigma_i$ can then be calculated as the total diffusive flux of
monomers into the boundary,
\begin{equation}
\sigma_i=2\pi RD\left({dP\over dr}\right)_{r=R}=2\pi D\rho 
\left(R\over{X_S}\right){K_1(R/X_S)\over K_0(R/X_S)}
\end{equation}
The cross section for monomer-monomer encounters $\sigma_o$ is
obtained from the same formula substituting $R$ by the monomer radius,
and $D$ by $2D$ as corresponds to relative diffusion.

The rate equations, with the explicit expressions for $\sigma_i$ and
$\sigma_o$ are still too complicated to solve exactly. At best we can
focus on extreme cases. The first step then is to evaluate the extreme
values of $\sigma_i$ for the case in which $R \gg X_S$ and $R \ll
X_S$. Using the known asymptotic values of the Bessel functions, and
once again omitting numerical constants, we have
\begin{equation}
\sigma_i\sim\cases{{D\rho R\over{X_S}} & for $R  \gg X_S$,\cr\cr
                   {D\rho\over \ln(X_S/R)} & for $R \ll X_S$.\cr}
\end{equation}
From here on we will neglect the logarithmic variations appearing in
the cross section. To evaluate $\sigma_o$ we notice that the monomer
radius is a small constant (it is usually taken as half the unit of
length), so for all values of $X_S$, to a good approximation we have
\begin{equation}
\sigma_o\sim D\rho
\end{equation}

Formally, the diffusive cross section for monomers should vary in the
case $X_S \ll 1$, but this is of no consequence since in that regime
the contributions from diffusion are negligible.

Now we turn to the evaluation of the direct impingement flux $J$. As
was the case above, the exact evaluation of $J$ is rather difficult,
as it results from the solution of a moving boundary problem. Thus,
once again, we must resort to a quasistatic approximation.

The flux $J$ will consist of two terms, $J_b$ the contribution due to
direct impingement at the exterior boundary of the island, i.e. {\it
on} the substrate, and $J_i$ the contribution due to particles falling
on the island, diffusing to the boundary, and ``falling'' over the
edge to increase the size of the island. We will continue to assume
that the islands are compact, thus, the contribution due to direct
impingement at the boundary will be given by $FR$ (where we have
omitted geometrical constants). To evaluate the second contribution
we must solve:
\begin{equation}
D^* \left({\partial^2 P\over \partial r^2} +{1\over r}{\partial P\over
 \partial r}\right)+F-{1\over\tau_e^*} P({\bf r})=0
\end{equation}
subject to the boundary condition $P(r=R)=0$. We recall that $D^*$ and
$\tau_e^*$ represent the values of the diffusion constant and
evaporation time of the particles on a substrate of their same species
(clearly, for homoepitaxy $D^*=D$ and $\tau_e^*=\tau_e$).  The
boundary condition corresponds to the simplest situation, in which
there is no barrier at the edge of the island and every particle
reaching the boundary falls. The increase in coverage will be given by
the total flux accross the barrier, i.e. $J_i\sim-R D^* {\partial
P\over \partial r}$.  In writing this equation we have assumed that no
nucleation occurs on top of the island, for this to be the case, we
must require that $P_{max} \ll 1$ on the island, which will be the
case for small enough $F$.

The solution to the above equation can be readily found, and from it
we find that
\begin{equation}
J_i\sim R X_s^* F \left( {I_1(R/X_s^*)\over I_0(R/X_s^*)} \right)
\end{equation}
where $X_s^*=\sqrt{D^* \tau_e^*}$ is the typical distance a monomer
can diffuse on an island before desorption, and $I_0(x)$ and $I_1(x)$
are modified Bessel functions.  Again using the known properties of
Bessel functions, we can distinguish two limiting behaviors:
\begin{equation}
J_i\sim \cases{{F R^2} & for ~$R \ll X_S^*$,\cr
                   {{F R X_S^*}} & for` $R \gg X_S^*$.\cr}
\end{equation}

\subsection{High Evaporation Regimes}

We are now in a position to analyse the limiting cases of the system
described by the rate equations. First we consider the new behaviour
brought about by the presence of evaporation. We define the high
evaporation regimes as the systems in which all but the first two
terms on the right-hand side of Eq.\ \ref{rho} are negligible. Then,
since the coverage is small, we have
\begin{equation}
{d\rho\over dt}\sim F-{\rho\over\tau_e},
\end{equation}
and $\rho$ reaches a steady state value $\rho=F\tau_e$ in a time of
order $\tau_e$. Thus, as mentioned above, the high evaporation regimes
are characterized by having a constant monomer density (after an
initial transient) throughout most of the evolution of the
system. Under these circumstances Eq.\ \ref{islands} becomes trivial
and predicts that the number of islands at time $t$ is given by
\begin{equation}
N \sim F^2 \tau_e \left[1+X_S^2\right]t
\end{equation}

If the evaporation rate is very high, so that $X_S \ll 1$, then even
the smallest islands at the earliest stages of evolution will satisfy
the relation $R \gg X_S$ (will refer to this situation as extreme
incomplete condensation).  If we assume that the diffusion length on
the islands $X_s^*$ is also much less than one, then Eq.\ \ref{cove}
becomes
\begin{equation}
{d\theta\over dt}\sim FRN \sim F(\theta N)^{1/2},
\end{equation}
where we have neglected the increase of coverage due to the formation
of small new islands in Eq. \ref{cove}, as most of the coverage is due
to the large islands. And the number of islands evolves as
\begin{equation}
N \sim F^2 \tau_e t,
\end{equation}
Solving for $\theta$, we obtain
\begin{equation}
\theta\sim F^4 \tau_e t^3.
\end{equation}
Now comes a crucial assumption, namely that coalescence occurs when
the coverage reaches a fixed constant value, say $\theta_c= 1/6$, and
that this value is essentially independent of the details of how the
surface is covered.  Thus, the time $t_c$ at which coalescence occurs
will be
\begin{equation}
t_c\sim \left(F^4 \tau_e \right)^{-1/3}
\label{ftcdir}
\end{equation}
Since the number of islands is a monotonically increasing function of
time up to the time of coalescence, we can estimate the maximum number of 
islands on the surface as
\begin{equation}
N_{max}\sim N(t_c)\sim \left(F\tau_e\right)^{2/3}
\end{equation}

If instead we consider the case in which $R\gg X_s^*\gg 1$, and $X_s\ll 1$,
then island formation will still only occur via direct impingement,
but the increase of coverage will be primarily due to monomers landing
in the capture zone {\it on} the islands. In this situation, we will
have
\begin{equation}
{d\theta\over dt}\sim X_S^* FRN,
\end{equation}
and again
\begin{equation}
N \sim F^2 \tau_e t,
\end{equation}
from which we obtain
\begin{equation}
\theta\sim F^4 \tau_e X_s^* t^3.
\end{equation}
The coalescence time can be estimated as before, leading to a maximum
number of islands that scales as
\begin{equation}
N_{max}\sim N(t_c)\sim \left(F\tau_e X_s^*\right)^{2/3}
\end{equation}
If we instead consider the case of extreme mismatch between the
materials, so $X_S\ll 1$ but $X_S^*\gg R$ throughout the evolution of the
system, the the capture zone on the islands becomes the whole island,
at which point the coverage increases as
\begin{equation}
{d\theta\over dt}\sim F \theta,
\end{equation}
where we used $NR^2\sim \theta$.  Coalescence then occurs at at a time
$t_c\sim 1/F$, and the maximum number of islands at that time will be
\begin{equation}
N_{max} \sim F\tau_e,
\end{equation}

Different and more interesting regimes exist for the situation in
which $X_S \gg 1$, still in the high evaporation regime, i.e. the
evolution of the monomer density is still essentially controlled
solely by the deposition and evaporation. This requirement actually
limits the value of $X_S$. The upper bound for the range of $X_S$ in
which this situation prevails is discussed below. First we consider
the case in which eventually $R \gg  X_S^*$ as well. In this situation,
the increase in coverage is due mainly to the particles deposited
within the capture zone of the islands, which we assume to be
relatively large.

For times long enough for these conditions to prevail, the evolution
of the coverage will be given by
\begin{equation}
{d\theta\over dt}\sim (X_S+X_S^*) R FN,
\end{equation}
and the number of islands will evolve as
\begin{equation}
{dN\over dt}\sim F^2 \tau_e X_S^2.
\end{equation}
Thus, solving for the coverage, we find that
\begin{equation}
\theta\sim F^4 X_S^2 (X_S+X_S^*)^2 \tau_e t^3.
\end{equation}
Once again we assume that coalescence becomes important at a value of
$\theta$ of the order of unity, so that the coalescence time in this
regime will be given by
\begin{equation}
t_c\sim (F^4 X_S^2(X_S+X_S^*)^2 \tau_e)^{-1/3},
\label{ftc1}
\end{equation}
and the maximum number of islands will be given by
\begin{equation}
N_{max}\sim \left({FX_S^2\tau_e\over X_S+X_S^*}\right)^{2/3}.
\label{nmaxdifev}
\end{equation}

Note that ignoring the particles that land on the islands, i.e. taking
$X_S^*=0$, as if there was an infinite edge berrier, yields the same
scaling as would be expected if $X_S\gg X_S^*$, and also the same
scaling as would be expected in homoepitaxy, where $X_S=X_S^*$. Thus,
ignoring these "second layer" particles is not crucial for the
calculation of the exponents, at least for a large range of the values
of the parameters.

If we now assume that eventually $R\gg X_S$, but that $X_S^*\gg R$, then
once again the coverage will evolve as
\begin{equation}
{d\theta\over dt}\sim F \theta.
\end{equation}
Coalescence then occurs at at a time $t_c\sim 1/F$, and the maximum number of 
islands at that time will now be given by
\begin{equation}
N_{max} \sim F\tau_e X_S^2
\end{equation}
This is the result obtained by Venables, as we discuss later (section
VI).

\subsection{Low evaporation rates}

If $\tau_e$ is further increased, then the system does not reach the
regime characterized in the previous section. As mentioned above, this
regime holds as long as $R \gg X_S$ beyond certain point in the
evolution of the system. We can estimate the value of $\tau_e$ at
which this condition never holds, i.e. when $X_S\sim R_{max}$, where
the maximum island size $R_{max}$ can be estimated through
\begin{equation}
R_{max}\sim \left({\theta_c\over N_{max}}\right)^{1/2}\sim
          \tau_e^{-1/2} F^{-1/3} D^{-1/6}
\end{equation}
Hence, the evaporation time above which the condition $R  \gg X_S$ never
holds is such that
\begin{equation}
X_S=\left({D \tau_e}\right)^{1/2}\sim R_{max}\sim \tau_e^{-1/2} F^{-1/3} 
D^{-1/6}
\end{equation}
that is,
\begin{equation}
\tau_e^{(2)}\sim (F D^2)^{-1/3}={\ell_{CC}}^2 \tau
\end{equation}
For evaporation times above $\tau_e^{(2)}$ the monomers are expected
to travel distances much longer than the largest typical island sizes.
At this point it is important to realize that our criterion for the
onset of coalescence, $\theta\sim 1$, is equivalent to the requirement
$R\sim \ell$, where $\ell\sim N^{-1/2}$ is the typical distance
between islands.  Thus, for evaporation times much larger than
$\tau_e^{(2)}$, a great part of the evolution of the system takes
place with the monomers traveling distances which are greater than the
inter-island distances. During this phase we expect the effects of
evaporation to be negligible. In effect, if $\tau_e>\tau_e^{(2)}$, the
kinetics of the monomer density is no longer determined solely by the
evaporation, but rather, it is eventually determined by the
aggregation processes.

 At very short times the number of islands is expected to be very
small. Therefore, in the early stages of evolution, they cannot affect
the monomer density. Thus, the early stages of evolution are expected
to be similar to those of the previous regime, for example:
\begin{equation}
\rho \sim F \tau_e, \qquad N \sim F^2 X_S^2 \tau_e t \qquad \theta 
\sim F^3 X_S^4 \tau_e t^2.
\end{equation}
This situation is expected to hold until the number of islands is
large enough, so that the term $\sigma_iN$ is no longer negligible
when compared to $F$. That is, until a time $t_{xx}$ such that
\begin{equation}
\sigma_iN\sim DF^3X_S^2\tau_e^2 t_{xx}\sim F,
\end{equation}
from which we obtain
\begin{equation}
t_{xx}\sim(F^2X_S^4\tau_e)^{-1}.
\end{equation}
As expected, the number of islands at $t_{xx}$ is $N(t_{xx})\sim
1/X_S^2$, indicating that evaporation effects ``turn'' off once the
typical inter-island distance is smaller than $X_S$. For times beyond
$t_{xx}$ all evaporation effects become negligible and our rate
equations reduce to those given by Tang \cite{tang} for a system
without evaporation (plus the direct impingement terms that are
negligible in this limit) for which the results are well known. Namely
the time at which coalescence occurs is given by
\begin{equation}
t_c\sim 1/F
\label{ftc2}
\end{equation}
and the maximum number of islands nucleated on the surface is
\begin{equation}
N_{max}\sim \left({F\over D}\right)^{1/3}
\end{equation}

We can now summarize now the various regimes obtained.  
There are three principal regimes which are spanned as the
evaporation time $\tau_e$ decreases. We call them, in this order, the
{\it complete condensation} regime where evaporation is not important,
the {\it diffusion} regime where islands grow mainly by diffusive
capture of monomers and finally the {\it direct impingement} regime
where evaporation is so important that islands can grow only by
capturing monomers directly from the vapor.  Within each of these
regimes, there are several subregimes characterized by the value of
$X_S^*$, the desorption length on top of the islands. 
We use $\ell_{CC} \equiv
(F \tau)^{-1/6}$, the typical island-island distance when there is no
evaporation and $R_{max}$ as the maximum island radius, reached
 at the onset of coalescence.

{\it complete \ condensation } $X_S \gg \ell_{CC}$

\begin{equation}
N_{max}\sim F^{1/3}\tau^{1/3} {\rm \ for \ any \ X_S^*}
\label{eqcc}
\end{equation}

\vspace{1cm}

{\it diffusive growth } $1 \ll X_S \ll \ell_{CC}$

\begin{equation}
N_{max}\sim\cases{(FX_S^2\tau_e)^{2/3}(X_S+X_S^*)^{-2/3} & {\rm if}~~~~
                   $X_S^* \ll R_{max}$ \  (a)\cr\cr
                   F\tau_eX_S^2 & {\rm if}~~~~ $X_S^* \gg R_{max}$ \  (b)\cr\cr}
\label{eqdif}
\end{equation}

with $R_{max}\sim (X_S+X_S^*)^{1/3}(FX_S^2\tau_e)^{-1/3}$, which gives for
the crossover between regimes (a) and (b) : $X_S^*(crossover) \sim (FX_S^2\tau_e)^{-1/2}$.

\vspace{1cm}

{\it direct \ impingement \ growth} $X_S \ll 1$

\begin{equation}
N_{max}\sim\cases{(F\tau_e)^{2/3} & {\rm if}~~~~$X_S^* \ll 1$ \ (a)\cr\cr
            ({F\tau_e})^{2/3}{X_S^*}^{-2/3}& {\rm if}~~~~$1 \ll X_S^* \ll R_{max}$ \ (b)\cr\cr
                   F\tau_e & {\rm if}~~~~ $X_S^* \gg R_{max}$ \  (c)\cr\cr}
\label{eqdir}
\end{equation}

with $R_{max}\sim ({F\tau_e})^{-1/3} {X_S^*}^{1/3}$, which gives for
the crossover between regimes (a) and (b) : $X_S^*(crossover) \sim (F\tau_e)^{-1/2}$.

We note that these equations agree with the scaling analysis presented
above (Equations \ \ref{scpinev} and \ref{scpiev}) in the appropriate
subregimes ($X_S^* \leq X_S$).

\section{Computer simulations}

In the following paragraphs, we test the assumptions and predictions
of the analysis given in the preceding sections in the special case
$i^*=1$ and no contribution from atoms deposited on top of the
islands. As we stressed above, the exponents observed without
contribution from the second layer are the same as those observed for
a large range of parameters. We also show results that are not
attainable from this mean-field calculations, namely the island size
distributions.

Our computer simulations generate sub-monolayer structures using the
four processes included in our model (see the introduction). Here we
take $\tau=1$ as the time scale of our problem. The monomer diffusion
coefficient is then given by $D=1/4$. We use triangular lattices (six
directions for diffusion) of sizes up to $2000 \times 2000$ with
periodic boundary conditions to avoid finite size effects. For
simplicity, in these simulations, the atoms deposited on top of
existing islands are not allowed to "fall" on the substrate. We have
checked that allowing the atoms to fall down the substrate does not
change significantly the results.

The program actually consists of a repeated loop. At each loop, we
calculate two quantities $p_{drop} = F / (F + \rho ({1\over \tau_e} +
{1\over \tau}))$ and $p_{dif} = (\rho /\tau) / (F + \rho ({1\over
\tau_e} + {1\over \tau}))$ that give the respective probabilities of
the three different processes which could happen : depositing a
particle (deposition), moving a particle (diffusion) or removing a
particle from the surface (evaporation). More precisely, at each loop
we throw a random number p ($0 < p < 1$) and compare it to $p_{drop}$
and $p_{dif}$.  If $p < p_{drop}$, we deposit a particle; if $p >
p_{drop} + p_{dif}$, we remove a monomer, otherwise we just move a
randomly chosen monomer.  After each of these possibilities, we check
whether an aggregation has taken place and go to the next loop (for
more details, see \cite{boston}).

\subsection{Checking the assumption : constant coverage for saturation}

\begin{figure}
\centerline{
\hbox{(a)
\epsfxsize=6cm
\epsfbox{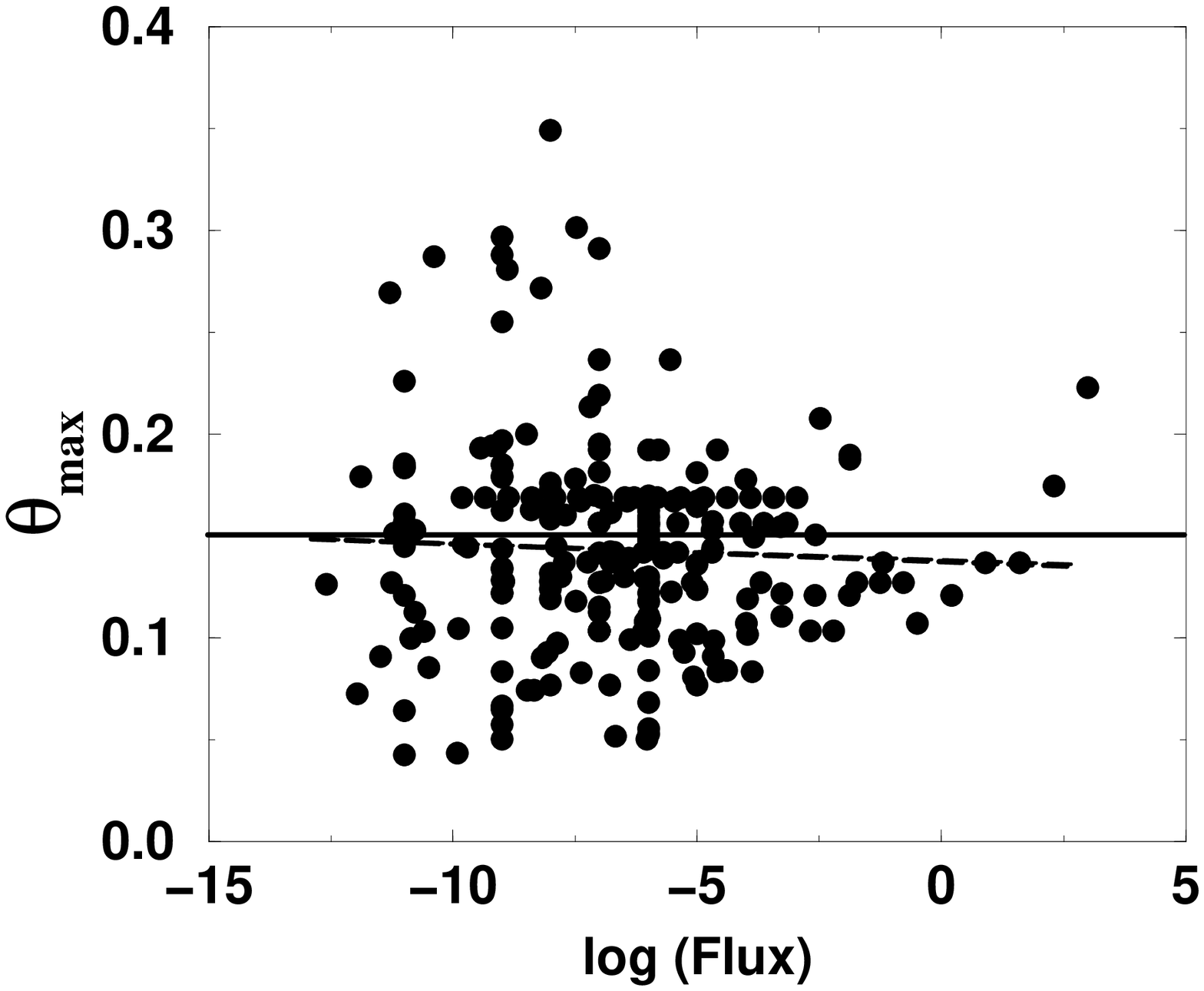}
}}
\centerline{
\hbox{(b)
\epsfxsize=6cm
\epsfbox{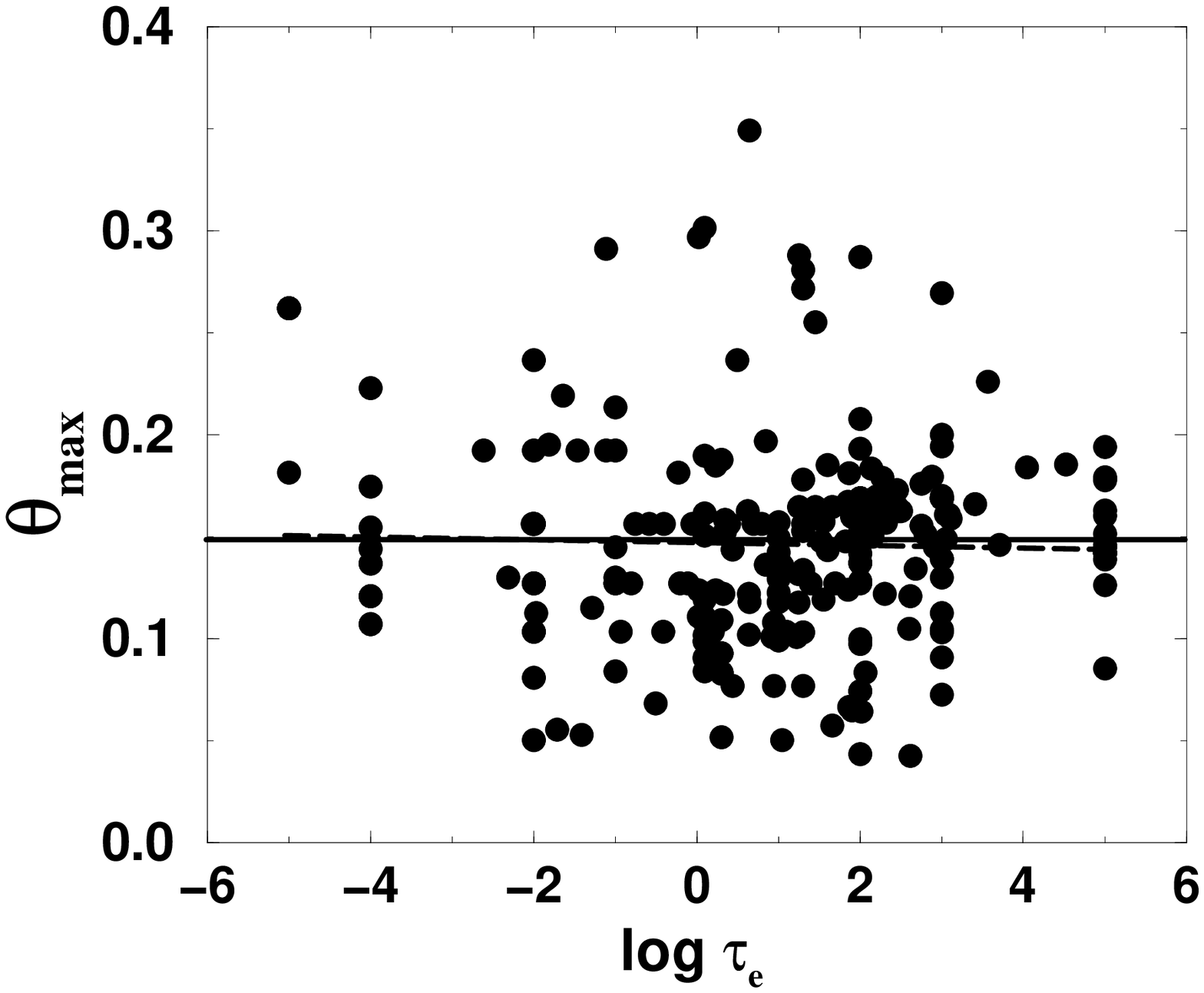}
}}
\centerline{
\hbox{(c)
\epsfxsize=6cm
\epsfbox{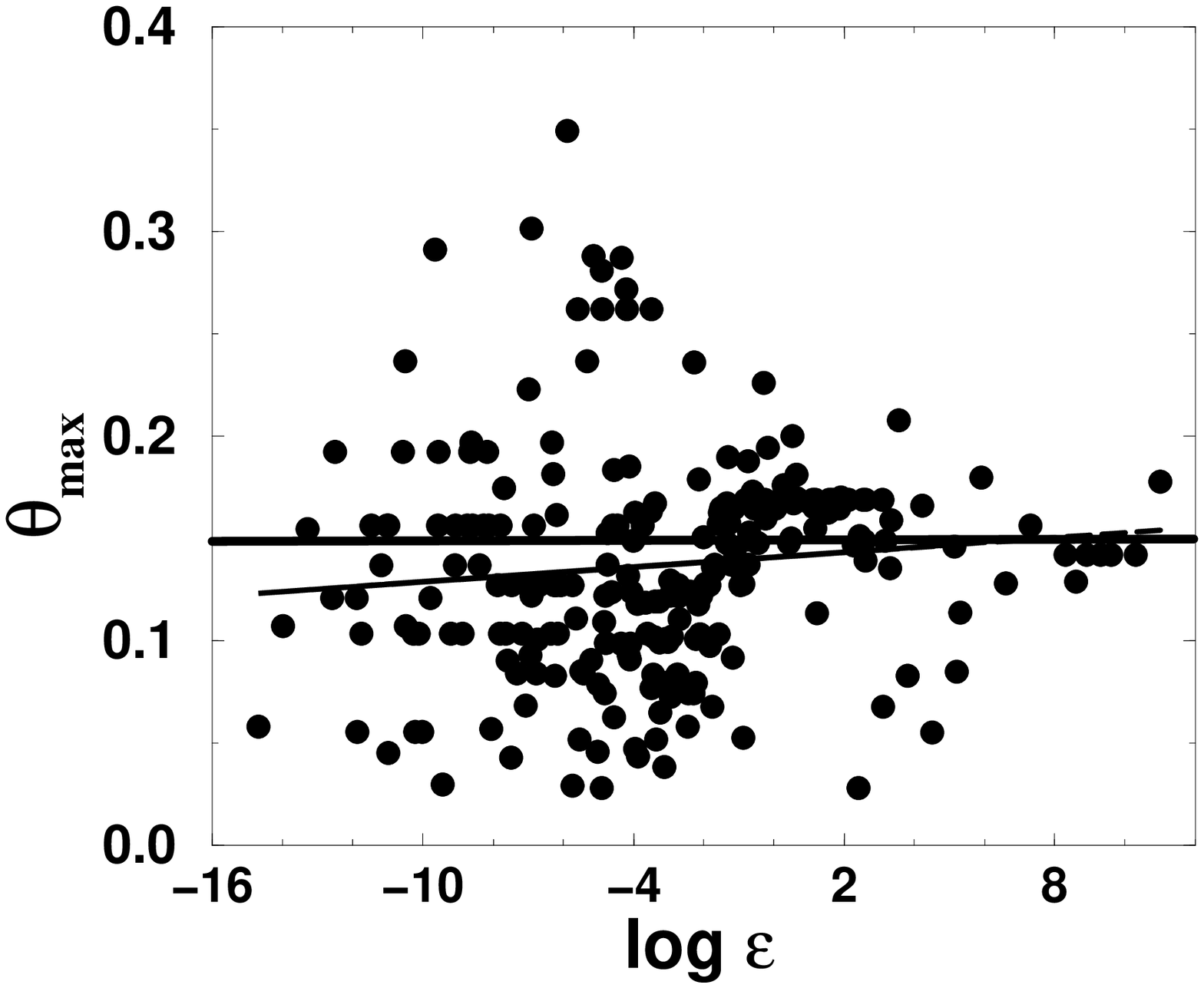}
}}
\caption{ 
Coverage ($\theta_{max}$) at which the island density
reaches its maximum. The different figures show the (non) evolution of
$\theta_{max}$ versus (a) the flux F (b) the evaporation time $\tau_e$
and (c) the evaporation parameter $\epsilon$. The solid lines indicate
our assumption, namely that the maximum island density is reached at a
constant coverage. The dashed lines show linear {\it fits} of the
data.  In all these graphs, there is no significative evolution of the
coverage at saturation as a function of the deposition parameters, as
can be seen in the coefficients of the fits : (a) $\theta_{max} = 0.13
- 0.00085 log(F)$; (b) $\theta_{max} = 0.15 - 0.00069 log(\tau_e)$;
(c) $\theta_{max} = 0.14 - 0.0012 log(\epsilon)$. We only notice an
increase of the fluctuations when the evaporation becomes important
since the systems contains only a few islands. These fluctuations are
due to the fact that each point in the figures represents a single
run, with no averaging.}
\label{covmax}
\end{figure}

A major assumption made in the theoretical treatment is that the
maximum of the island density is reached when the {\it coverage}
attains a constant value. We note that this assumption is equivalent
to the one used in the scaling analysis, since, for compact islands,
$\theta \sim N \pi R^2 \sim (R/l)^2$.  It is essential then to check
this assumption first. Figs \ \ref{covmax}a-c show that this it is
justified, since $\theta_{max}$, the coverage at which the maximum
island density is reached, does not vary systematically with F,
$\tau_e$ or $\epsilon$. $\epsilon$ is defined as ${\left(1+
X_S\right)}X_S^{5}\left(\frac FD\right)$ and indicates the importance
of evaporation : $\epsilon \ll 1$ means that evaporation is
significant, while $\epsilon \gg 1$ indicates that we are in the
regime of complete condensation (see Eq. \ \ref{eqsca}).

In the non-evaporation case, it has also been recognized that
$\theta_{max}$ is independent on the flux or diffusion rate
\cite{villain,tang,model}. Actually, Villain and co-workers
\cite{villain} use this criterion to find the flux and diffusion
dependence of $N_{max}$. We think that the combination of two effects
can lead to a constant $\theta_{max}$. First, at coverages close to
.15, islands occupy enough surface to capture rapidly the landing
monomers, which prevents nucleation of new islands. Second, at this
coverage, islands begin to touch and coalesce ($R \sim \ell$), thus
starting the decrease in island density. We note that $\theta_{max}$
is {\it not} constant when large islands are allowed to move, even in
the non-evaporation case \cite{boston}.

Having confirmed our main assumption, we now turn on to the evolution
of the maximum island density $N_{max}$ as a function of the different
parameters $F$ and $\tau_e$ (remember that we take $\tau=1$ as the
time scale of our problem).

\subsection{Checking the crossover scaling}

\begin{figure}
\centerline{
\hbox{
\epsfxsize=10cm
\epsfbox{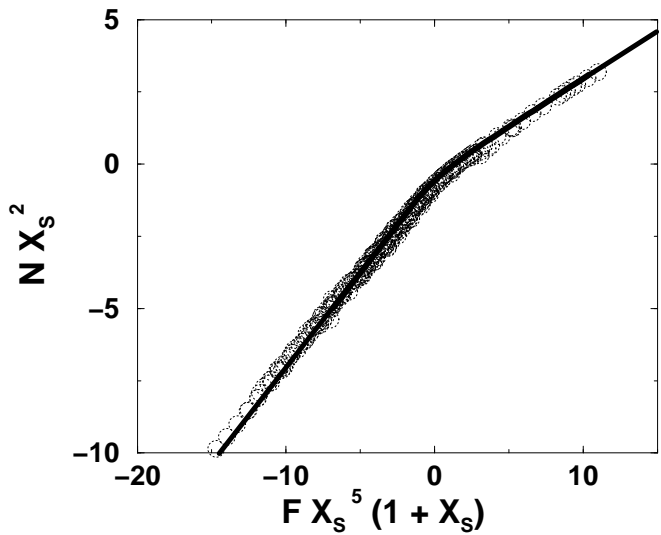}
}}
\caption{
 Universal function rescaling all our data. As predicted by
Equation \ \protect\ref{eqsca}, the normalized island density $N_{max}
X_S^2$ follows a single curve as a function of the evaporation
parameter $\epsilon = {F \over D} {X_S}^5 (1+ X_S)$.  The solid curve
shows the function predicted in the text (Equation \
\protect\ref{function}), while the circles represent the results of
the computer simulations.}
\label{crossover}
\end{figure}

Before looking in detail into the different regimes predicted by
Equations \ \ref{eqcc}, \ref{eqdif} and \ref{eqdir},
we summarize our simulation results in
Fig. \ \ref{crossover}. We show there {\it all} our data (more than
200 points) for $N_{max}$ as a function of the parameters. Our scaling
analysis predicts that the data should fall into a single curve, given
by Equation \ \ref{eqsca}. We see that the data remarkably confirms
our analysis, over more than 20 orders of magnitude. This gives us
confidence on our entire approach and its predicted exponents, which
we now turn on to check in more detail.

\subsection{Checking the exponents}

The object of this section is to check that the results summarized in
Equations \ \ref{eqcc}, \ref{eqdif} and \ref{eqdir} are correct.

\subsubsection{
Scaling of the maximum island density as a function of incident flux}

\begin{figure}
\centerline{
\hbox{
\epsfxsize=10cm
\epsfbox{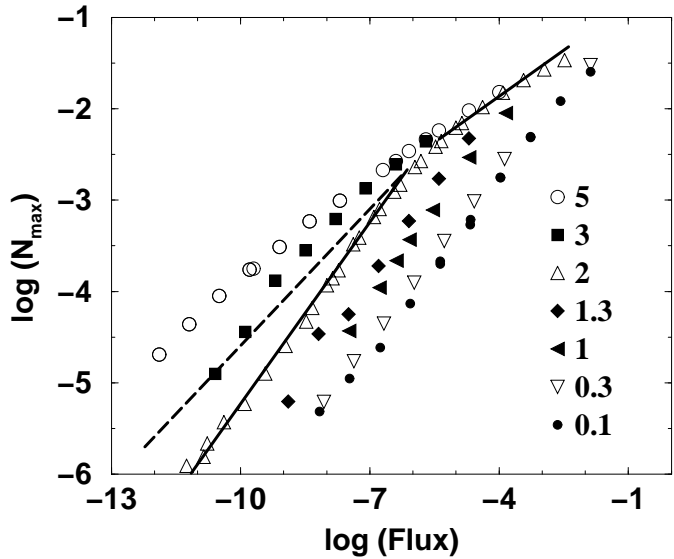}
}}
\caption{
Evolution of the maximum island density as a function of the
flux for different evaporation times. The number next to each symbol
corresponds to the log($\tau_e$) value for that set (remember that
$\tau \equiv 1$). The solid lines show the expected values for the
exponents : 2/3 for low values of the flux (evaporation is
significant) and 1/3 for higher values (complete condensation).  The
dashed line shows the prediction of other authors ($N_{max} \sim
F^{1/2}$ for low fluxes).  See the discussion (section VI) for
details.}

\label{nmaxfnoscal}
\end{figure}

Figure \ \ref{nmaxfnoscal} shows the evolution of the maximum island
density as a function of the flux for different evaporation times.
Each of these curves is different from the others, since they
correspond to different evaporation times. However, according to our
preceding analysis, they should all present a transition from the low
evaporation regime to the high evaporation regime.  This can be
detected by a change of slope, from $N_{max} \sim F^{2/3}$ in the high
evaporation regime to $N_{max} \sim F^{1/3}$ in the low one (Eq. \
\ref{eqdif}a and \ref{eqcc}).  
Of course, this regime change does not occur for all
the curves at the same value of the Flux, since the parameter that
determines that change is not the Flux but rather $X_S^2 =
\tau_e/\tau$.  Figure \ \ref{nmaxfnoscal} shows our predictions are
correct, at least concerning the Flux evolution of the maximum island 
density. We now turn to the other variable, the evaporation time.

\subsubsection{Maximum island density as a function of evaporation time}

\begin{figure}
\centerline{
\hbox{
\epsfxsize=10cm
\epsfbox{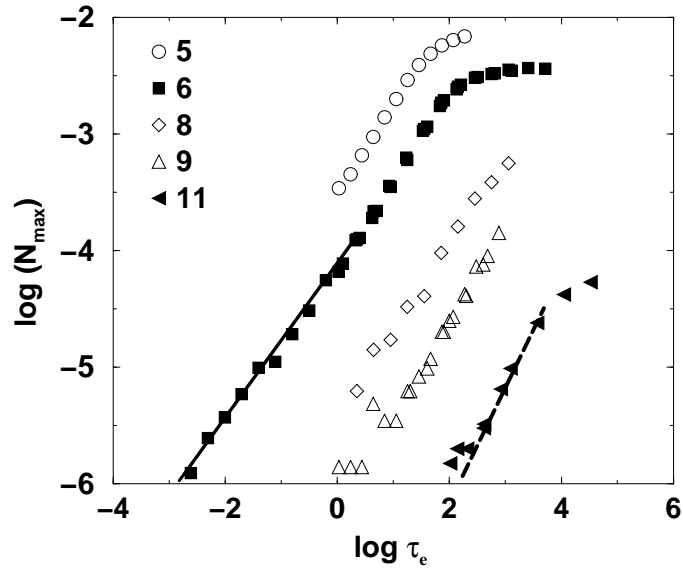}
}}
\caption{
Maximum island density as a function of the evaporation time
for different fluxes. The number next to each symbol corresponds to
the log(1/F) value for that set. The solid line shows the expected
value for the exponent : 2/3 for low values of $\tau_e$ (evaporation
is significant). The dashed line shows the exponent expected for the
intermediate regime.}
\label{nmaxtenoscal}
\end{figure}

We show in figure \ \ref{nmaxtenoscal} the dependence of the maximum
island density on $\tau_e$. We notice that for high enough evaporation
times, the island density tends to become roughly constant, as
predicted by our calculations.  For lower values of $\tau_e$,
$N_{max}$ changes rapidly. Our analysis predicts two regimes :
for $1 \ll \tau_e \ll F^{-1/3}$, we expect $N_{max} \sim \tau_e$,
(Eq. \ \ref{eqdif}a) while for $\tau_e \ll 1$, we expect $N_{max} \sim
{\tau_e}^{2/3}$ (Eq. \ \ref{eqdir}a).
This last regime is clearly seen for the curve
obtained for a flux $F=10^{-6}$ (squares, solid line).  The
intermediate regime is difficult to see because of the crossovers with
the two other regimes. However, taking a very low value for the flux
($F=10^{-11}$, filled triangles), we can see that the slope in this
intermediate regime is close to 1 (dashed line).

\subsubsection{Direct impingement regime}

\begin{figure}
\centerline{
\hbox{
\epsfxsize=10cm
\epsfbox{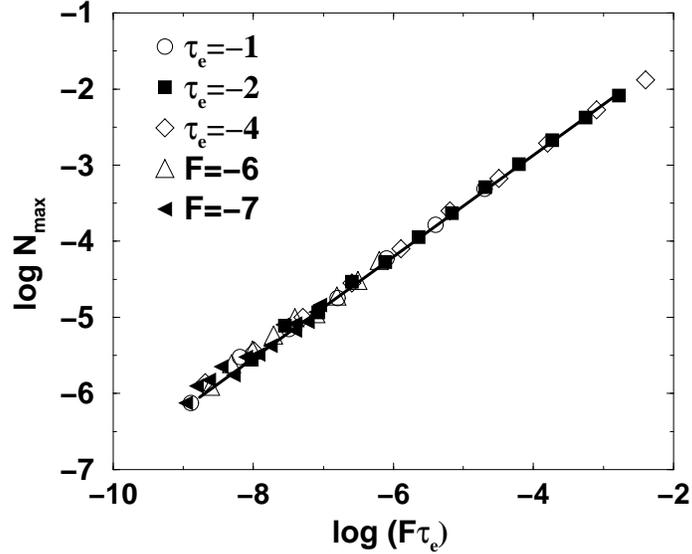}
}}
\caption{
Dependence of the maximum island density $N_{max}$ on $F
\tau_e$ in the direct impingement regime ($X_S \ll 1$). The solid line shows our prediction ($N_{max} \sim (F
\tau_e)^{2/3}$ for extremely low evaporation times, Eq. \
\ref{eqdir}a). We see that our prediction remarquably fits the data
over more than 6 orders of magnitude.  The numbers indicate the values
of logarithms of the evaporation time $\tau_e$ or the flux $F$ for the
different sets of data.}
\label{fte23}
\end{figure}

We have also checked the exponents obtained when $X_S \ll 1$, in the
direct impingement regime. Eq. \ \ref{eqdir}a predicts that the
maximum island density scales with the product $F \tau_e$ with an
exponent 2/3. This is confirmed by our computer simulations (Figure \
\ref{fte23}), over more than 6 orders of magnitude.

\subsubsection{Condensation coefficient at the maximum island density}

\begin{figure}
\centerline{
\hbox{
\epsfxsize=10cm
\epsfbox{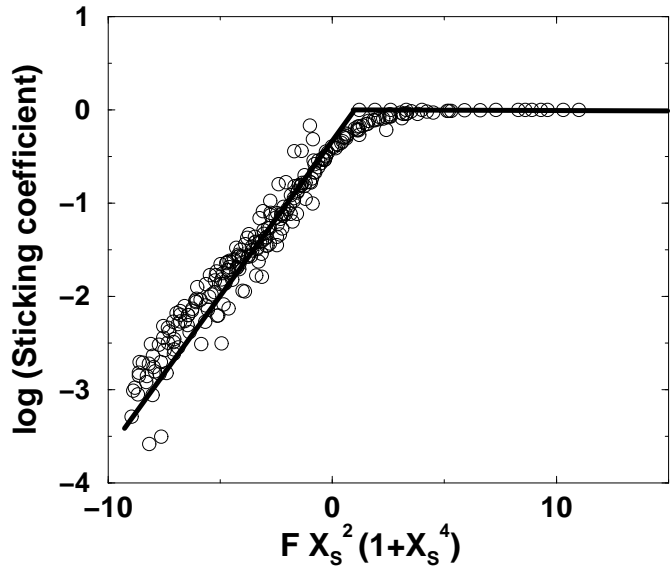}
}}
\caption{
 Evolution of the sticking coefficient at the maximum island
density as a function of the evaporation parameter $F X_S^2 (1+
X_S^4)$. The solid lines represent our predictions (Eqs. \
\protect\ref{ftc1} and \ \protect\ref{ftc2}).  See the text for
details.}
\label{stick}
\end{figure}

A last test for the analysis is presented in Fig \ \ref{stick}.  We
show that the dependence of the sticking coefficient S at the
saturation island density ($S = \theta_{max} / Ft_c$) follows Eqs. \
\ref{ftc1}, \ref{ftcdir} and \ \ref{ftc2}.  By constructing an
evaporation parameter $\eta = F X_S^2 (1+ X_S^4)$, we can group all
the regimes in a single curve : for the complete condensation regime,
$S \sim \eta^0 \sim 1$, for the others $S \sim \eta^{1/3}$.

\subsection{Island size distributions}

\begin{figure}
\centerline{
\hbox{
\epsfxsize=10cm
\epsfbox{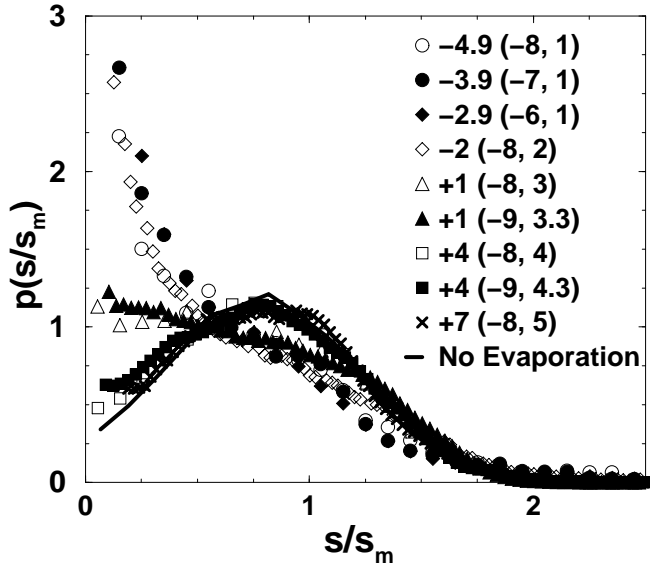}
}}
\caption{
 Rescaled island size distributions for different values of
the evaporation parameter $\epsilon$ and different coverages $\theta$
between .05 and .2. $p(s/s_m)$ represents the probability to find an
island containing s monomers when the average size is $s_m =
\theta/N$. If $n_s$ is the number of islands containing s monomers, we
have $p(s/s_m) = s_m * n_s / N_t$ where $N_t$ is the total number of
islands.  The solid line shows the size distribution obtained without
evaporation.  The first number next to each symbol corresponds to log
$\epsilon$. The two other figures correspond to log $F$ and log
$\tau_e$ respectively ($\tau=1$).}
\label{distrib}
\end{figure}

Island size distributions have proven very useful as a tool for
experimentalists to distinguish between different growth mechanisms
\cite{bardotti,stroscio}. By {\it size} of an island, we mean its
total number of monomers or mass. Unfortunately this information is
beyond the reach of the simple mean field rate equation analysis
presented above \cite{bales}. Nevertheless, the distributions can be
obtained from the simulations. Fig \ \ref{distrib} shows the evolution
of the {\it rescaled} \cite{model,stroscio} island size distributions
as a function of the evaporation parameter $\epsilon = {\left(1+
X_S\right)}X_S^{5}\left(\frac FD\right)$.  It is clear that the
distributions are significantly affected by the evaporation, smaller
islands becoming more numerous when evaporation increases. This trend
can be qualitatively understood by noting that new islands are created
continuously when evaporation is present, while nucleation rapidly
becomes negligible in the complete condensation regime. The reason is
that islands are created (spatially) homogeneously in the last case,
because the positions of the islands are correlated (through monomer
diffusion), leaving virtually no room for further nucleation once a
small portion of the surface is covered ($\theta \sim 0.05$). In the
limit of strong evaporation, islands are nucleated randomly on the
surface, the fluctuations leaving large regions of the surface
uncovered. These large regions can host new islands even for
relatively large coverages, which explains that there is a large
proportion of small ($s < s_m$) islands in this regime.

\section{Discussion}

Other authors have analyzed similar mean-field rate equations to find
the growth dynamics and maximum island density in the presence of
evaporation \cite{stowell,venables73,venables84,stoyanov}.
In what follows, we discuss the relationship between our work and the
preceding studies. In fact, a distinction must
be made between the result of Stoyanov and Kaschiev, and that of
Venables et al.

Stoyanov and Kaschiev consider that atoms may diffuse on top of an
island, with the same diffusion length as on the
substrate ($X_S = X_S^*$ in our notations). This corresponds to the
regimes described by Equations \ \ref{eqcc} and \ \ref{eqdif}a (they do
not consider the direct impingement regime). However, their result in the
regime where evaporation is important is $N_{max} \sim (F\tau_e)^{1/2}$,
different from our predictions (see both Eq. \ \ref{eqdif}a and 
Eq. \ \ref{eqdir}b), which are clearly supported by our simulations 
(see for example Fig. \ \ref{nmaxfnoscal}). As shown in the following, 
we think that the difference with our result
stems from overlooking the fact that only the adatoms falling at a
distance $X_s$ from the island edge contribute to the rate of growth
of the island : in other terms, their capture cross section is not
appropriate.

Stoyanov and Kaschiev assume, as we do, that the diffusing adatoms
have two possible fates: either to evaporate or to be captured by an
island. Their lifetime $\tau_l$ is thus
\begin{equation}
\frac1{\tau_l} = \frac1{\tau_e} + DN \;,
\end{equation} 
and the adatom density reads
\begin{equation}
\label{ada}
\rho\approx F\tau_l\approx \frac{F\tau_e}{1+D\tau_e N}\;.
\end{equation} 
This is Eq. (6.2) of \cite{stoyanov}.

At this point, the incorrect assumption is made that the rate of
growth of the area of an island of radius $R$ reads $d{R^2}/d t
\approx D\rho\approx FD\tau_e/(1+D\tau_e N)$ throughout the strong
evaporation regime (cf. Sec IV).  Writing then
\begin{equation}
\label{scall}
D\rho^{(i^*+1)} \approx \frac{d R^2}{dt} N^2 \approx D\rho N^2 \;,
\end{equation}
one finds $\rho^{i^*}\approx N^2$, or
\begin{equation}
 N^2(1+X_s^2 N)^{i^*}  \approx \left(\frac FD X_s^2\right)^{i^*} \;.
\end{equation}
This is Eq. (6.12) of \cite{stoyanov}, and it leads to $N\approx
(F/D)^{i^*/2}X_s^{i^*}$ in the ``incomplete condensation'',
$X_s^2N\ll1$.

On the other hand, Venables et al. actually study (though it may not
be easily guessed by reading their papers : we would like here
to acknowledge an anonymous referee for precious help) the regime where 
{\it all} the particles deposited on top of the islands {\it immediately} fall
on its border, thus contributing to increase its radius. This corresponds
to the subregime described by Eq. \ \ref{eqdif}b. Our prediction agrees 
with their result in the "Extreme Incomplete" regime (see Table 1 of Ref. 
\cite{venables84}). This regime applies only if diffusion
and/or desorption are different on the substrate and on top of
islands, which cannot be the case in homoepitaxial situations. In
heteroepitaxial growth, on the other hand, the substrate and the
islands of the first monolayer are chemically different, and Venables
et al.'s assumption may apply; however, it is a special situation, not
the general rule. The origin of their "initially incomplete" regime
is more mysterious. We try in the following to understand its origin.

We now show that the assumption of Venables et al.'s concerning
the atoms falling on top of an island, leads to their equation for the
maximum island density, Eq. (2.17) of Ref. \cite{venables84}. In our
notations the latter reads
\begin{equation}
\label{venn}
N(1+X_s^2N)^{i^*}(\theta_0+X_s^2N)\approx (F/D)^{i^*}X_s^{4i^*}\;.
\end{equation}
where $\theta=R^2N$ is the deposited dose (coverage) and $\theta_0$
its value at coalescence.

Assume that an island grows by capturing all atoms falling on top of
it, plus those diffusing to it. Then, the rate of growth of the island
area is
\begin{equation} 
\label{rate}
\frac{d R^2}{d t}\approx FR^2+D\rho\approx F\theta/N+D\rho\;,
\end{equation}
where $\theta=R^2N$ is the deposited dose, or coverage.

From Eq. (\ref{ada}) and the nucleation rate, one finds
\begin{equation}
\label{this}
\frac{1}{\tau_{\rm nuc}} \approx  D\rho^{i^*+1} \approx
D\left(\frac{F\tau_e}{1+D\tau_e N}\right)^{i^*+1}\;.
\end{equation}
 
Letting (\ref{this}) equal to $(d R^2/d t)N^2$ (cf. Eq. \ref{scall}),
one has
\begin{equation}
\label{scal111}
D\left(\frac{F\tau_e}{1+D\tau_e
N}\right)^{i^*+1} \approx 
F\theta_0N+D\frac{F\tau_e}{1+D\tau_e N}N^2\;,
\end{equation}
where $\theta_0<1$ is the coverage at coalescence. Rewriting yields
\begin{equation}
\label{scalll}
\frac{(F/D)^{i^*}X_s^{4i^*}}{(1+X_s^2 N)^{i^*}} \approx 
\left[\theta_0(1+X_s^2 N)+X_s^2N\right]N\approx (\theta_0+X_s^2N)N
\;,
\end{equation}
which gives immediately Eq. (\ref{venn}). 

We now turn on the relation of our crossover scaling (section III-C)
to recent work by Bales \cite{balle}.
Recently, the crossover scaling of the island density between
different growth regimes {\it on vicinal surfaces} (without
evaporation) was addressed by Bales
\cite{balle}, and by Pimpinelli and Peyla \cite{pipe}. In particular,
Bales claims that ``when atoms desorb from the surface a crossover
scaling form identical to'' his results ``is satisfied with'' the
distance between surface steps ``replaced by the average distance
$X_s$ a monomer will travel before desorbing'' \cite{balle}.

We argue that Bales' claim is not true, as it stands. To see this, we
need consider what happens on a vicinal substrate. We do this using
the argument of Sect. III. Of course, competition between two
length scales occurs in this case, too, the two lengths being the
adatom diffusion length before capture by an island, $\ell_s$, and the
step-step distance $\ell$. When $\ell_s\ll\ell$, the diffusing adatoms
behave as though they were on a flat substrate, and the island density
is still given by equation (4). In the opposite case, $\ell_s\gg\ell$,
the adatom density is fixed by capture at steps, so that $\rho$ can be
found by replacing $X_s$ by $\ell$ in Eq. (5), in agreement with
Bales' statement:
\begin{equation}
\label{balle1}
\rho\approx F\ell^2/D \;.
\end{equation}

However, since evaporation is now neglected, an island grows by
capturing all adatoms diffusing to its border. This means the island
average radius grows as
\begin{equation}
R\approx \ell \sqrt{Ft}\;,
\end{equation}
and since $R\approx \ell_s$ at $t=t_c$, one has
\begin{equation}
\frac 1{t_c}\approx F\frac{\ell^2}{\ell_s^2}\approx F\ell^2N \;.
\end{equation}

In the regime of dominant adatom capture by steps the island density
at saturation is found from
\begin{equation}
\label{balle2}
D\rho^{i^*+1}\approx  \frac{N}{t_c}
\end{equation}
and thus
\begin{equation}
\label{pippi}
{N}\approx \left(\frac FD \ell^2 \right)^{i^*/2}\;.
\end{equation}

A crossover between equation (4) and (\ref{pippi}) is thus expected:
it has been actually found by Pimpinelli and Peyla \cite{pipe}.

Bales \cite{balle} considered the crossover scaling of the island
density at low coverage. His result for the regime of dominant adatom
capture by steps, Eq. (9) of his paper, is found (within logs) by
letting $t_c= t=\theta/F$ and $i^*=1$ in (\ref{balle2}) above,
\begin{equation}
\label{pippi2}
{N}\approx \frac FD \ell^4 \theta \;.
\end{equation}

Note that this cannot be obtained by replacing $X_s$ by $\ell$ in our
result, Eq.  (7) above. It is however obtained by replacing $X_s$ by
$\ell$ in Venables' result. This is because Venables does not take
into account the crucial phenomenon which occurs during
evaporation--namely, that only the adatoms falling near or at the step
edge can contribute to the island growth. To neglect this is correct
within Venables' assumptions, as well as on a vicinal surface without
evaporation, where the latter mechanism is replaced by adatom capture
by steps.

Our crossover scaling does not therefore coincide with Bales', and
Bales' results \cite{balle} however modified do not describe
homoepitaxial growth with desorption ($X_S = X_S^*$).

\section{Summary and Perspectives}

By combining different mean-field analysis and extensive computer
simulations, we have shown that the presence of evaporation has
important effects on the growth of submonolayer films. We have
investigated the different regimes that arise when the growth
parameters are varied, and predicted the behaviour of several
experimentally accessible quantities such as the island size
distributions, the maximum island density and the time at which this
maximum is reached. In some cases, by measuring these last two
quantities we can infer the values of both the evaporation and the
diffusion times, which are difficult to obtain otherwise. For example,
if the experiments are carried in the intermediate regime (and this
can be checked from the sticking coefficient), by using 
Equations \  \ref{eqdif}a  and  \ \ref{ftc1}, 
we obtain $\tau = F^2 \left( t_c N_{max} \right)^3$ and $\tau_e = t_c 
{N_{max}}^2$.

 We have also shown that our model is more general than previous studies
of growth in presence of evaporation.

Future directions of study should include more realistic hypotheses
for a direct comparison with experiments \cite{henry}. In particular,
we are presently extending our analysis to the case of
three-dimensional growth (i.e. leading to cap-shaped islands) and the
influence of defects on the surface which could act as nucleation
centers.

We wish to thank M. Meunier and C. Henry for helpful discussions, Ph.
Nozi\`eres for a critical reading of the manuscript and an anonymous referee
for precious comments on the connection of our analysis with previous
works.

e-mail addresses : jensen@dpm.univ-lyon1.fr,
hernan@ce.ifisicam.unam.mx, pimpinelli@ill.fr

\end{document}